\pdfoutput=1
\documentclass[a4paper]{article}
\usepackage{color}
\usepackage{amsfonts,amssymb,graphicx,latexsym,url,diagrams}
\usepackage{amsmath}
\usepackage{bussproofs}
\usepackage{fullpage}
\usepackage{theorem}

{\theorembodyfont{\rmfamily}
\newtheorem{definition}{Definition}[section]}
{\theorembodyfont{\rmfamily}
\newtheorem{example}{Example}[section]}
{\theorembodyfont{\rmfamily}
\newtheorem{remark}{Remark}}
\usepackage[mathscr]{eucal}
\renewcommand{\theenumi}{\roman{enumi}}

\makeatletter
\renewcommand\p@enumii{(\theenumi)}
\makeatother
\newsavebox{\proofbox}
\savebox{\proofbox}{%
  \begin{picture}(7,7)\put(0,0){\framebox(7,7){}}\end{picture}%
}

\begin{document}
\title{Fuzzy Categories\\
       \colorbox{yellow}{Work in Progress!}}
\author{Apostolos Syropoulos\\
        Greek Molecular Computing Group\\
        366, 28th October Str.\\
        GR-671\ 00\ \ Xanthi, Greece\\
        \texttt{asyropoulos@yahoo.com}}
\date{Autumn 2014}
\maketitle
\begin{abstract}
Since categories are graphs with additional ``structure'', one should start from
fuzzy graphs in order to define a theory of fuzzy categories. Thus is makes
sense to introduce categories whose morphisms are associated with a plausibility 
degree that determines to what extend it is possible to ``go'' from one
object to another one. These categories are called {\em fuzzy categories}.
Of course, the basic properties of these categories are similar but not
identical to their ordinary counterparts. Thus, it is necessary to 
introduce notion like fuzzy commutative diagrams, fuzzy initial and fuzzy
terminal objects, etc. 
\end{abstract}
\section{Introduction}
Categories, which were invented by Samuel Eilenberg and Saunders Mac Lane, form a very 
high-level abstract mathematical theory that unifies all branches of mathematics. The 
standard reference on category theory is Mac Lane's book-length introduction to
categories~\cite{mclane98}. Category theory plays a central role in modern mathematics and 
theoretical computer science, and, in addition,  it is used in mathematical 
physics (e.g., see~\cite{flori10}), in  software engineering~\cite{fiadeiro04}, etc. 
However, what makes  category theory even more  interesting is that it is an alternative to set 
theory as a foundation  for mathematics. Indeed, as Mac~Lane~\cite{mclane86} pointed out
\begin{quote}
It is now possible to develop almost all of ordinary Mathematics in a well-pointed topos
[i.e., an ordinary category that has some additional properties] with choice and a natural 
number object. The development would seem unfamiliar; it has nowhere been carried out yet in 
great detail. However, this possibility does demonstrate one point of philosophical interest: 
The foundation of Mathematics on the basis of set theory (ZFC) is by no means the only 
possible one!
\end{quote}

Roughly, a category is a universe that includes all mathematical objects of a particular form 
together with maps between them. These maps must obey a few basic principles. For example, the 
collection of all sets and functions between them forms the category that is traditionally 
denoted by $\mathbf{Set}$. Obviously, one can form categories of fuzzy structures.
For instance, Carol Walker~\cite{walker04} and Siegfried Gottwald~\cite{gottwald06} have 
defined such categories. However, some other categories of fuzzy structures emerged from efforts 
to define fuzzy models of linear logic. In particular, Michael Barr~\cite{barr91}, Basil Papadopoulos 
and this author~\cite{pap-syr00,pap-syr05,syropoulos06} have introduced such categories. 

Not so surprisingly, a category of fuzzy structures is not a fuzzy structure itself. Indeed,
Alexander \v{S}ostak~\cite{sostak99} was the first researcher who had realized this. In order to
remedy this situation, \v{S}ostak introduced a new structure that mimics the way fuzzy 
sets are defined as is evident from the following definition:
\begin{definition}
An L-fuzzy category (where $L$ is a GL-monoid) is a quintuple $\mathcal{C} = (\mathrm{Ob}(\mathcal{C}),\omega, 
M(\mathcal{C}), \mu, \circ)$ where $\mathcal{C}_{\bot} = (\mathrm{Ob}(\mathcal{C}), M(\mathcal{C}), 
\circ)$ is a usual (classical) category called the bottom frame of the fuzzy category $\mathcal{C}$; 
$\omega : \mathrm{Ob}(\mathcal{C}) \rightarrow L$ is an $L$-subclass of 
the class of objects $\mathrm{Ob}(\mathcal{C})$ of $\mathcal{C}_{\bot}$ and 
$\mu:M(\mathcal{C})\rightarrow L$ is an $L$-subclass of the class of morphisms 
$M(\mathcal{C})$ of $\mathcal{C}_{\bot}$ . Besides $\omega$ and $\mu$ must satisfy the following conditions:
\begin{enumerate}
\item  if $f : X \rightarrow Y$ , then $\mu(f ) \le\omega(X) \wedge ω(Y)$;
\item  $\mu(g \circ f )\ge \mu(g) \ast \mu(f)$ whenever composition $g\circ f$ is defined,
where $\ast$ is a binary operator that obeys a number of rules;
\item if $e_{X} : X \rightarrow X$ is the identity morphism, then $\mu(e_{X}) = \omega(X)$.
\end{enumerate}
\end{definition}
For reasons that will become clear later on, \v{S}ostak's approach is not the ideal
solution to the problem of the ``fuzzification'' of category theory. Instead, by following
a different path one should be to define an alternative form of fuzzy categories. In particular, 
since any category can be identified with a graph (the inverse is not true), I am using the 
results of fuzzy graph theory to define, what I think are, real fuzzy categories. 
This approach is justified by the fact that (meta)categories are introduced using notions from (meta)graph theory. 
Interestingly, G{\'e}rard Huet and Amokrane Sa{\"\i}bi~\cite{huet00} have followed a similar line 
of thought in order to define constructive categories and categorical structures.

One may wonder why someone should get into the trouble to define fuzzy categories. Indeed,
this is a quite reasonable question since there is no room for more meaningless generalizations.
However, if fuzziness, in particular, and {\em vagueness}, in general, are fundamental properties
of this cosmos, then we should use this fact to define deviant Mathematics.\footnote{Obviously,
according to quantum mechanics our world is vague, but probabilistic systems do not fundamentally 
differ from their deterministic counterparts (e.g., nondeterministic and deterministic Turing 
machines have exactly the same computational power). Nevertheless, it seems that this does not 
apply to fuzziness.} Thus, if this is true, then we could use fuzzy categories as a tool to
develop the foundations for this deviant Mathematics.

\section{Basic Ideas and Concepts}
As was explained in the introduction, I will define fuzzy categories using known notions
from fuzzy graph theory. Thus, it is important to recall the notions that are necessary to
define fuzzy categories. Let us start with fuzzy graphs:
\begin{definition}
A {\em fuzzy graph}~\cite{lee05} consists of a collection of nodes and a collection of arrows 
between these nodes. Each arrow must have a specific {\em domain} node (i.e., its source), 
a {\em codomain} node (i.e., its target), and a plausibility degree, which expresses the grade 
to which it is possible to go from the domain to the codomain of an arrow. 
\end{definition}
Obviously, it is possible to have two or more different arrows that have the same domain 
and codomain, but, clearly, different plausibility degrees. The notation 
``$f:A\overset{\rho}{\rightarrow}B$'' means that $f$ is an arrow that goes from $A$ (the domain) 
to $B$ (the codomain) with plausibility degree that is equal to $\rho\in[0,1]$. Alternatively, 
one can use the following notation:
\begin{displaymath}
A\overset{f}{\underset{\rho}{\longrightarrow}}B.
\end{displaymath}
\begin{definition}
Let $k>0$. In a fuzzy graph $\mathcal{G}$ a path from a node  $X$ to a node $\Psi$ of length
$k$ is a sequence $(f_1,f_2,\ldots,f_k)$ of arrows, which are not necessarily distinct, such that
\begin{displaymath}
X\overset{f_k}{\underset{\rho_k}{\longrightarrow}} A_{k-1}
\overset{f_{k-1}}{\underset{\rho_{k-1}}{\longrightarrow}}\ldots A_{1}
\overset{f_{1}}{\underset{\rho_{1}}{\longrightarrow}} \Psi
\end{displaymath} 
and the plausibility degree of the path is $\min_{i=1}^{k}\rho_{i}$.
\end{definition}
\begin{remark}
We can use any t-conorm, but for reasons of simplicity we use min.
\end{remark}
Having defined fuzzy arrows and fuzzy arrow composition, we can proceed with the definition 
of fuzzy categories.
\begin{definition}
A fuzzy category $\mathscr{C}$ comprices
\begin{enumerate}
\item a collection of entities called {\em objects};
\item a collection of entities called {\em arrows} or {\em morphisms};
\item operations assigning to each $\mathscr{C}$-arrow $f$ a $\mathscr{C}$-object 
$A=\mathop{\mathrm{dom}} f$, its domain, a $\mathscr{C}$-object $B=\mathop{\mathrm{cod}} f$, 
its codomain, and a plausibility degree 
$\rho=\mathop{\mathrm{p}} f$.  Typically, the plausibility degree is a real number 
belonging to the unit interval. These operations on $f$ are indicated by displaying $f$ as an arrow
starting from $A$ and ending at $B$ with plausibility degree $\rho$:   
\begin{displaymath}
A\overset{f}{\underset{\rho}{\longrightarrow}}B\quad\mbox{or}\quad
f:A\overset{\rho}{\rightarrow}B;
\end{displaymath}
\item an operation assigning to each pair $(g,f)$ of arrows with $\mathop{\mathrm{dom}} g=
\mathop{\mathrm{cod}} f$, an aror $g\circ f$, the {\em composite} of $f$ and $g$, having
$\mathop{\mathrm{dom}}(g\circ f)=\mathop{\mathrm{dom}} f$,
$\mathop{\mathrm{cod}}(g\circ f)=\mathop{\mathrm{cod}} g$, and 
$\mathop{\mathrm{p}}(g\circ f)=\min\{\mathop{\mathrm{p}} f, \mathop{\mathrm{p}} g\}$. This
operation and the previous three are subject to the {\em associative law}: Given the
configuration
\begin{displaymath}
A\overset{f}{\underset{\rho_1}{\longrightarrow}}B
\overset{g}{\underset{\rho_2}{\longrightarrow}}C
\overset{h}{\underset{\rho_3}{\longrightarrow}}D
\end{displaymath}
of  $\mathscr{C}$-objects and $\mathscr{C}$-arrows, then $h\circ(g\circ f)=(h\circ g)\circ f$;
\item an {\em assignment} to each $\mathscr{C}$-object $B$ of a $\mathscr{C}$-arrow
$\mathbf{1}_{B}:B\overset{1}{\rightarrow}B$, called the {\em identity arrow on} $B$, such
that the following {\em identity law} holds true:
\begin{displaymath}
\mathbf{1}_B\circ f=f\quad\mbox{and}\quad g\circ\mathbf{1}_B=g
\end{displaymath} 
for any $\mathscr{C}$-arrows $f:A\overset{\rho_f}{\rightarrow}B$ and
$g:B\overset{\rho_g}{\rightarrow}A$.
\end{enumerate}
\end{definition}
\begin{remark}
Obviously, every ordinary category is a fuzzy category with arrows that have plausibility
degree equal to 1.
\end{remark}
\begin{example}
Let us give a relatively simple example of a fuzzy category that has as objects sets.
Assume that $f:X\rightarrow Y$ is function. Then we say that $f$ is an
arrow from $X$ to $Y$ with plausibility degree $\lambda$ if there are fuzzy subsets that
are characterized by the functions $A:X\rightarrow\mathrm{I}$
and $B:Y\rightarrow\mathrm{I}$ such that $B(f(x))-A(x)\ge\lambda$, for all $x\in X$. Assume that
$f:X\overset{\lambda_1}{\rightarrow}Y$ and $g:Y\overset{\lambda_2}{\rightarrow}Z$ are two
arrows. Then since $B(f(x))-A(x)\ge\lambda_1$ and $C(g(f(x)))-B(f(x))\ge\lambda_2$, for all $x\in X$, 
one concludes that $C(g(f(x)))-A(x)\ge\lambda_3$, for all $x\in X$ and where 
$\lambda_3=\min\{\lambda_1,\lambda_2\}$. Thus, the
composite arrow $g\circ f:X\overset{\lambda_3}{\rightarrow}Z$ exists and can be defined
form its constituents. It is not difficult to see that the associative law holds also. Clearly, 
the identity arrow for some object $X$ is the identity function of this set. Also, it 
is almost trivial to see that the identity law holds. The resulting fuzzy category 
will be called \textbf{FSet}.
\end{example}
\begin{remark}
The objects of a fuzzy category can be fuzzy structures, but this is something that should
not concerns us. After all, category theory is about arrows and their properties and
not about objects.
\end{remark}
\begin{example}
Assume that $R:X\times X\rightarrow[0,1]$ is a fuzzy relation such that 
$R(x,x)=1$ for all $x\in X$ and $R(x,y)\mathbin{\ast}R(y,z)\le R(x,z)$ for
all $x,y,z\in X$ and where $\ast$ is a t-norm. A fuzzy relation with these properties
is called a {\em $\ast$-fuzzy preorder}. When the t-norm is function $\min$, then it will
be called just fuzzy preorder. Any fuzzy preorder relation $R$ determines a fuzzy preorder 
category $P$ in which the arrows $p\overset{\rho}{\rightarrow}p'$ are exactly those pairs 
$\langle p,p'\rangle$ for which $R(p,p')=\rho$. The fuzzy relation is transitive, which 
implies that there is a unique way of composing arrows. Also, the fuzzy relation is 
reflexive and so there are the necessary identity arrows. 
\end{example}
\begin{example}
Let $\wedge$ be the only object of a category and let us identify all arrows of this category
with their plausibility degrees. For example $\wedge
\overset{\lambda}{\underset{\lambda}{\longrightarrow}}\wedge$
is the arrow $\lambda$ whose plausibility degree is obviously $\lambda$. Given two arrows
$\lambda_1$ and $\lambda_2$, and assuming that $\wedge$ denotes the minimum, as is usually the
case, then $\lambda_1\circ\lambda_2=\lambda_1\wedge\lambda_2$. Assume there is an arrow 
$1_{\wedge}$ such that $1_{\wedge}\circ\lambda=\lambda$ and $\lambda\circ1_{\wedge}=\lambda$
for all arrows $\lambda$, then according to the definition of arrow composition this arrow
is the identity arrow, that is, $1_{\wedge}=1$. Note that it is quite possible to have more than
one arrow that has plausibility degree equal to one, nevertheless, for our purposes these
arrows will behave exactly like $1$ does. In different words, they will be isomorphic,
but we will say more about isomorphisms in a while.
\end{example}
\begin{example}
A category is a {\em deductive system} (for example, see~\cite{lambek94} for a thorough
description of this categories-as-deductive-systems paradigm). In this paradigm, objects
are seen as {\em formulas}, arrows as {\em proofs} (or deductions), and an operation 
on arrows as a {\em rule of inference}. In particular, each arrow $f:A\rightarrow B$ is 
thought of as the ``reason'' why $A$ entails $B$. Thus, the identity law is the reason
why $A$ entails $A$, for all $A$ objects (formulas) and the associative law becomes the
following rule of inference:
\begin{prooftree}
\AxiomC{$f:A\overset{}{\longrightarrow}B$}
\AxiomC{$g:B\overset{}{\longrightarrow}C$}
\BinaryInfC{$f\circ g:A\overset{}{\longrightarrow}C$}
\end{prooftree}
Similarly, a fuzzy category is a {\em fuzzy deductive system} in which objects may be 
{\em fuzzy formulas} (remember, the objects of any fuzzy category are not necessarily ``crisp''), 
arrows are {\em fuzzy deductions}, and the associative law is the following fuzzy inference:
\begin{prooftree}
\AxiomC{$f:A\overset{\rho_f}{\longrightarrow}B$}
\AxiomC{$g:B\overset{\rho_g}{\longrightarrow}C$}
\BinaryInfC{$f\circ g:A\overset{\rho_{f\circ g}}{\longrightarrow}C$}
\end{prooftree}
\end{example}

\paragraph{Commutative diagrams} In category theory {\em commutative diagrams} play the role
equations play in algebra. In the simplest case a commutative diagram can be identified with 
two different paths starting from the same object $A$ and ending with the same object $B$ in 
which the composition of the arrows that make up the first path and the composition of the arrows of 
the second path yield two arrows that have the same effect (i.e., when applied to the same
object(s), they yield the same result). In general, when dealing with fuzzy arrows we need
to stick to this requirement, but we distinguish at least two different cases. In the first
case, the plausibilities must be exactly the same while in the second case, they must be
greater than a specific minimum. In particular, assume that
\begin{align*}
A&\overset{f_n}{\underset{\lambda_n}{\longrightarrow}}\ldots
\overset{f_{1}}{\underset{\lambda_{1}}{\longrightarrow}}B\\
A&\overset{g_m}{\underset{\rho_m}{\longrightarrow}}\ldots
\overset{g_{1}}{\underset{\rho_{1}}{\longrightarrow}}B
\end{align*} 
are two paths. Then these paths form a {\em strong} commutative diagram provided that
\begin{displaymath}
\min\Bigl\{\lambda_1,\ldots, \lambda_n \Bigr\}=\min\Bigl\{\rho_1,\ldots, \rho_m \Bigr\}.
\end{displaymath}
Otherwise, we say that the two paths commute with plausibility degree $\nu$, where
\begin{displaymath}
\nu=\min\Bigl\{\min\{\lambda_1,\ldots,\lambda_n\},\min\{\rho_1,\ldots,\rho_m\}\Bigr\}.
\end{displaymath}
\begin{example}
The identity law can be expressed with the following strong commutative diagrams:
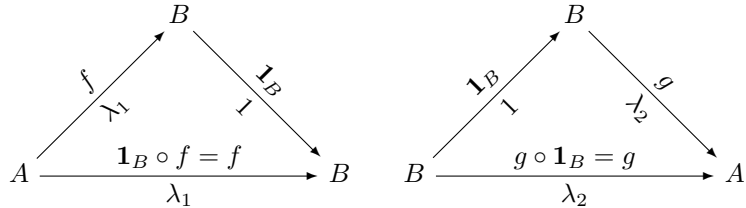
\begin{center}
\begin{tabular}{lcr} 
\begin{diagram}
& & B \\
& \ruTo^f_{\lambda_1} &  & \rdTo_1^{\mathbf{1}_B} \\
A & & \rTo_{\lambda_1}^{\mathbf{1}_{B}\circ f=f} & & B\\
\end{diagram} & \qquad &
\begin{diagram}
& & B \\
& \ruTo_1^{\mathbf{1}_B} &  & \rdTo^g_{\lambda_2}\\
B & & \rTo_{\lambda_2}^{g\circ\mathbf{1}_{B}=g} & & A\\
\end{diagram}
\end{tabular}
\end{center}
Obviously, $\lambda_1\le1$ and $\lambda_2\le1$.
\end{example}

\paragraph{Isomorphisms} In ordinary mathematics two entities of the same kind can be
isomorphic or not. In the fuzzy setting, they can be isomorphic up to some degree and
the following definition follows this principle:
\begin{definition}
Two objects $A$ and $B$ are isomorphic to some degree
$\lambda$ if there are arrows $f:A\overset{\lambda_1}{\longrightarrow}B$ 
and $g:B\overset{\lambda_2}{\longrightarrow}A$ such that the following diagrams
\begin{center}
\begin{tabular}{lcr} 
\begin{diagram}
& & B \\
& \ruTo^f_{\lambda_1} &  & \rdTo^g_{\lambda_2} \\
A & & \rTo_{1}^{\mathbf{1}_{A}} & & A\\
\end{diagram} & \qquad &
\begin{diagram}
& & A \\
& \ruTo^g_{\lambda_2} &  & \rdTo^f_{\lambda_1}\\
B & & \rTo_{1}^{\mathbf{1}_{B}} & & B\\
\end{diagram}
\end{tabular}
\end{center}
are commutative with degree that is equal to $\lambda$, where 
$\lambda=\min\{\lambda_1,\lambda_2\}$.
\end{definition}
Arrows can be {\em monic} or {\em epic} up to some degree. In particular, an arrow  
$f:A\overset{\lambda_1}{\longrightarrow}B$ is monic up to $\nu$ if there are two arrows
$g:C\overset{\lambda_2}{\longrightarrow}A$ $h:C\overset{\lambda_3}{\longrightarrow}A$
such that the following diagram
\begin{diagram}
C                    & \rTo^{g}_{\lambda_2} & A\\
\dTo_{\lambda_3}^{h} &                      & \dTo_{\lambda_1}^{f}\\
A                    & \rTo^{f}_{\lambda_1} & B
\end{diagram}
commutes with degree equal to $\nu=\min\{\lambda_1,\lambda_2,\lambda_3\}$. In
addition, $g\mathrel{=_{\nu}}h$, that is, $g$ and $h$ are equal with degree $\nu$.
Similarly, an arrow $f':A\overset{\kappa_1}{\longrightarrow}B$ is epic up to $\nu'$ 
if there are two arrows $g':B\overset{\kappa_2}{\longrightarrow}C$ $
h:B\overset{\kappa_3}{\longrightarrow}C$
such that the following diagram
\begin{diagram}
A                    & \rTo^{f'}_{\kappa_1} & B\\
\dTo_{\kappa_1}^{f'} &                      & \dTo_{\kappa_2}^{g'}\\
B                    & \rTo^{f}_{\kappa_3} & C
\end{diagram}
commutes with degree equal to $\nu'=\min\{\kappa_1,\kappa_2,\kappa_3\}$. In
addition, $g'\mathrel{=_{\nu'}}h'$, that is, $g'$ and $h'$ are equal with degree $\nu'$.
\paragraph{Initial and Terminal Objects} 
An object $T$ of a fuzzy category $\mathscr{C}$ is called terminal if
there is exactly one arrow $A\overset{1}{\longrightarrow}T$ for each object $A$ of
$\mathscr{C}$. An object of a fuzzy category that has a unique arrow with plausibility
degree equal to one {\em to} each object (including itself), is called an {\em initial object}. 

\section{Conclusions}
I have introduced fuzzy categories and some fuzzy categorical structures. There is
much work ahead! First one needs to define fuzzy functors, then fuzzy natural
transformations. However, the most important of all is to see whether these categories
have interesting properties and whether they can be used to solve interesting problems.


\begin{thebibliography}{10}

\bibitem{barr91}
Michael Barr.
\newblock Fuzzy models of linear logic.
\newblock {\em {Mathematical Structures in Computer Science}}, 6(3):301--312,
  1996.

\bibitem{fiadeiro04}
Jos{\'e}~Luiz Fiadeiro.
\newblock {\em {Categories for Software Engineering}}.
\newblock Springer-Verlag, New York, 2004.

\bibitem{flori10}
Cecilia Flori.
\newblock {A topos formulation of history quantum theory}.
\newblock {\em {Journal of Mathematical Physics}}, 51:053527--1--053527--31,
  2010.

\bibitem{gottwald06}
Siegfried Gottwald.
\newblock {Universes of Fuzzy Sets and Axiomatizations of Fuzzy Set Theory.
  Part II: Category Theoretic Approaches}.
\newblock {\em {Studia Logica}}, 84:23--50, 2006.

\bibitem{huet00}
G{\'e}rard Huet and Amokrane Sa{\"\i}bi.
\newblock {Constructive Category Theory}.
\newblock In {\em {Proof, language, and interaction: essays in honour of Robin
  Milner}}, pages 239--275. {The MIT Press}, Cambridge, MA, USA, 2000.

\bibitem{lambek94}
Joachim Lambek and Philip~J. Scott.
\newblock {\em Introduction to higher order categorical logic}.
\newblock {Cambridge University Press}, Cambridge, U.K., 1994.

\bibitem{mclane86}
Saunders~Mac Lane.
\newblock {\em {Mathematics: Form and Function}}.
\newblock Springer-Verlag, New York, 1986.

\bibitem{mclane98}
Saunders~Mac Lane.
\newblock {\em {Category Theory for the Working Mathematician}}.
\newblock Springer-Verlag, New York, second edition, 1998.

\bibitem{lee05}
Kwang~H. Lee.
\newblock {\em {First Course on Fuzzy Theory and Applications}}.
\newblock Springer-Verlag, Berlin, 2005.

\bibitem{pap-syr00}
Basil~K. Papadopoulos and Apostolos Syropoulos.
\newblock {Fuzzy Sets amd Fuzzy Relational Structures as Chu Spaces}.
\newblock {\em {International Journal of Uncertainty, Fuzziness and
  {K}nowledge-Based Systems}}, 8(4):471--479, 2000.

\bibitem{pap-syr05}
Basil~K. Papadopoulos and Apostolos Syropoulos.
\newblock {Categorical relationships between Goguen sets and ``two-sided''
  categorical models of linear logic}.
\newblock {\em {Fuzzy Sets and Systems}}, 149:501--508, 2005.

\bibitem{sostak99}
Alexander {\v{S}}ostak.
\newblock Fuzzy categories related to algebra and topology.
\newblock {\em Tatra Mountains Mathematical Publications}, 16(1):159--185,
  1999.
\newblock {Available from the web site of the Slovak Academy of Sciences}.

\bibitem{syropoulos06}
Apostolos Syropoulos.
\newblock {Yet Another Fuzzy Model for Linear Logic}.
\newblock {\em {International Journal of Uncertainty, Fuzziness and
  {K}nowledge-Based Systems}}, 14(1):131--135, 2006.

\bibitem{walker04}
Carol~L. Walker.
\newblock Categories of fuzzy sets.
\newblock {\em {Soft Computing}}, 8(4):299--304, 2004.

\end{thebibliography}
\end{document}